\documentclass[A4 paper,12pt,table]{article}

\title{\textbf{Identifying the Geometry of an Object Using Lock-In Thermography}}
\author{\textbf{BY:}\\ \\Xiao Tian and Meng Yuan Yin\\ \\National Junior College, Singapore\\ \\ \\\textbf{External Mentor}\\ \\Kok Hin Henry Goh\\ \\Email: henrygohkh@hotmail.com}
\date{}

\usepackage[top=1.7cm,bottom=1.7cm,left=1.7cm,right=1.7cm]{geometry}
\usepackage{amsmath} 
\usepackage{graphicx} 
\usepackage{float} 
\usepackage{subcaption} 
\usepackage{ulem} 
\usepackage[colorlinks=true,linkcolor=black, citecolor=blue]{hyperref} 

\DeclareMathOperator{\taninv}{tan^{-1}}

\newcommand{\HighlightBad}[1]{\textcolor{red}{#1}}
\newcommand{\HighlightCyan}[1]{\textcolor{cyan}{#1}}

\begin{document}


\maketitle

This report was written jointly by students (Tian and Yin) after an exploratory project aim at pushing the limits of Infrared Thermography. The report shows the following findings:
\begin{itemize}
\item{Results from Lock-In Thermography measurements show that it is possible to detect signals and features deeper than expected from conventional Lock-In Thermography}
\item{It is possible to automate the detection of geometries of relatively thick objects with relatively high success rates, even with non-uniform heating from halogen lamps}
\item{The techniques derived in the current work show that it is possible to detect geometries even with differential boundary effects such as additional radiation losses at the edges of geometries}
\item{From the current findings, it may be possible to derive techniques for real-time identification of geometries and embedded defects using Lock-In Thermography and cheap, portable equipment}
\end{itemize}


%
%
%
%

\newpage

\begin{abstract}

Lock-in Thermography (LIT) is a type of Infrared Thermography (IRT) that can be used as a useful non-destructive testing (NDT) technique for the detection of subsurface anomalies in objects. Currently, LIT fails to estimate the thickness at a point on the tested object. This makes LIT unable to figure out the 3-dimensional geometry of an object. In this project, two techniques of identifying the geometry of an object using LIT are discussed. The main idea of both techniques is to find a relationship between the parameters obtained from LIT and the thickness at each data point. Technique \text{I} builds a numerical function that models the relationship between thickness, Lock-In phase, and other parameters. The function is then inverted for thickness estimation. Technique \text{II} is a quantitative method, in which a database is created with six dimensions - thickness, Lock-In phase, Lock-In amplitude and three other parameters, based on data obtained from LIT experiments or simulations. Estimated thickness is obtained by retrieving data from the database. The database can be improved based on Principal Component Analysis. Evaluation of the techniques is done by measuring root-mean-square deviation, and calculating successful rate with different tolerances. Moreover, during the application of the techniques, Stochastic Gradient Descent can be used to determine the time when sufficient data have been collected from LIT measurement to generate the estimated geometry accurately.

\end{abstract}

\section{Introduction}

Infrared thermography (IRT) is a non-destructive testing (NDT) technique to detect shapes and anomalies of samples based on objects radiation in infrared range. It has many applications such as electrical, mechanical and insulation maintenance \cite{Bagavathiappan2013IPT}, the assessment of the performance and conditions of buildings \cite{Grinzato2002JCH}, the failure analysis of integrated circuits \cite{Schmidt2012MSE}, and the inspection of solar cells in renewable energy industry \cite{Breitenstein2003PIP}. IRT has two categories - passive thermography and active thermography. Passive thermography tests structures which are naturally at different temperature from the ambient \cite{Theodorakeas2015JOP}, while active thermography requires an external stimulus to induce relevant thermal contrast \cite{IbarraCastanedo2015UAM} into the tested object.

Lock-In Thermography (LIT) is one type of active thermography \cite{Maldague1996JAP}. During the process of LIT, a periodic heat wave is injected into the tested object. The injected thermal wave will cause the temperature at each pixel on the surface of the tested sample to vary over time. Simultaneously, the temperatures are recorded down by an infrared camera as raw thermal signals \cite{Sakagami2002IPT}. Variation in physical properties of an object such as thickness, thermal emissivity and existence of subsurface anomalies can lead to difference in the raw thermal signals at each pixel. The locations, sizes and shapes of anomalies of geometries, can be identified by analysing the Lock-In phase and amplitude calculated from the raw thermal signal curve. However, thickness of the objects has not been found to have a direct correlation to the signals. 
    
There are attempts made to derive the thickness of an object and the thickness of coating from the signal obtained from LIT. Coating thickness is predicted by observing and analysing the phase angle image of LIT \cite{Wu1996RPQ}. Quantitative evaluation of material thickness has been done by many research groups \cite{Christian2010}, \cite{Takahide20022}, \cite{Wallbrink2007}.  However, among the attempts, the tested samples are in steps or in constant gradients. This means that the conclusion of the current work done is restricted to particular situations as the gradient factor is not considered. Therefore, in this project, two techniques are developed to map the thickness to the thermal signal obtained from LIT. The two techniques can be adapted to objects with various gradients and are thus more accurate in thickness estimation. Besides, we can also use stochastic gradient descent (SGD) to generate the estimated geometry of the tested object simultaneously with the LIT measurement so as to increase the applicability of this project.

\section{Materials and Methods}

\subsection{Conducting LIT experiments and simulations}

Both experiments and simulations were conducted in this project. In order to ensure that the results are not restricted, we used FreeCAD to design 14 samples with different thickness and different types and values of gradients. Details of the samples are shown in Appendix~\ref{sec:Details of samples}.
 
To conduct experiments, the samples were printed out using Acrylonitrile Butadiene Styrene (ABS) in a 3D-printing machine. LIT experiments were then conducted in a setup as shown in Appendix~\ref{sec:Setup}. Samples were heated by a pair of halogen lamps with a Lock-In frequency of 0.01Hz. The raw thermal signals were captured by an infrared camera of 320$\times$240 pixels in the speed of 1 frame per second and transmitted to the laptop for further processing. Each sample was heated over a period of 200 seconds so that 201 frames were captured in total for each sample. 

Simulations of the experiment were also conducted in this project. This is because for LIT experiments, not all the radiation received by the infrared camera comes from the emission of the tested sample. It also comes from the emission of the surroundings and reflected by the object and the emission of the atmosphere \cite{Usamentiaga2014SS}. Moreover, since heat can be lost to the surroundings through convection, LIT experiments are affected by working conditions, such as the surrounding temperature, airflow and humidity. Therefore, there is a lot of thermographic data noise for experiments data \cite{Usamentiaga2008JOE}. Thus, simulations, which do not have data noise and are hence able generate clearer data, are also conducted to serve as a reference to benefit real-life work. The same 14 samples are imported into ElmerGUI to run simulations. The setup of simulations are the same as experiments so as to control the variables.

\subsection{Data processing}

After the raw thermal signal of the data points at each pixel was obtained, Lock-In calculations were applied in order to obtain the Lock-In phase and amplitude, which are two useful parameters in Lock-In studies. The equations used for Lock-In calculations are attached in Appendix~\ref{sec:Lock-In calculation} \cite{Goh2018arXiv}. Besides, by plotting thermal signal against time for each pixel, We noticed that all the graphs are similar. A typical thermal signal plot for a single pixel over the measurement period of $200$~s is shown in Appendix~\ref{sec:signal}. 

Other than the Lock-In amplitude and phase, a set of other parameters were derived to pinpoint the thickness of the sample at every pixel more accurately. All these parameters were stored for every relevant image pixel on the samples for each measurement, and subsequently used to generate datasets. 

\subsection{Technique \text{I} - Building a numerical function}

Despite that currently there is no theoretical function that relates the parameters to thickness, we are able to fit the relationship to known functions by observing its trend \cite{Harvey2004}. Hence, the first technique that we discuss is to build a universal function that models the relationship between the parameters and thickness. We plotted graphs of every parameter against each other, and found that one of the parameters, Parameter $\mu$, and phase have clear relationship.  

\begin{figure}[H]
\centering
\includegraphics[width = 110mm, height = 65mm]{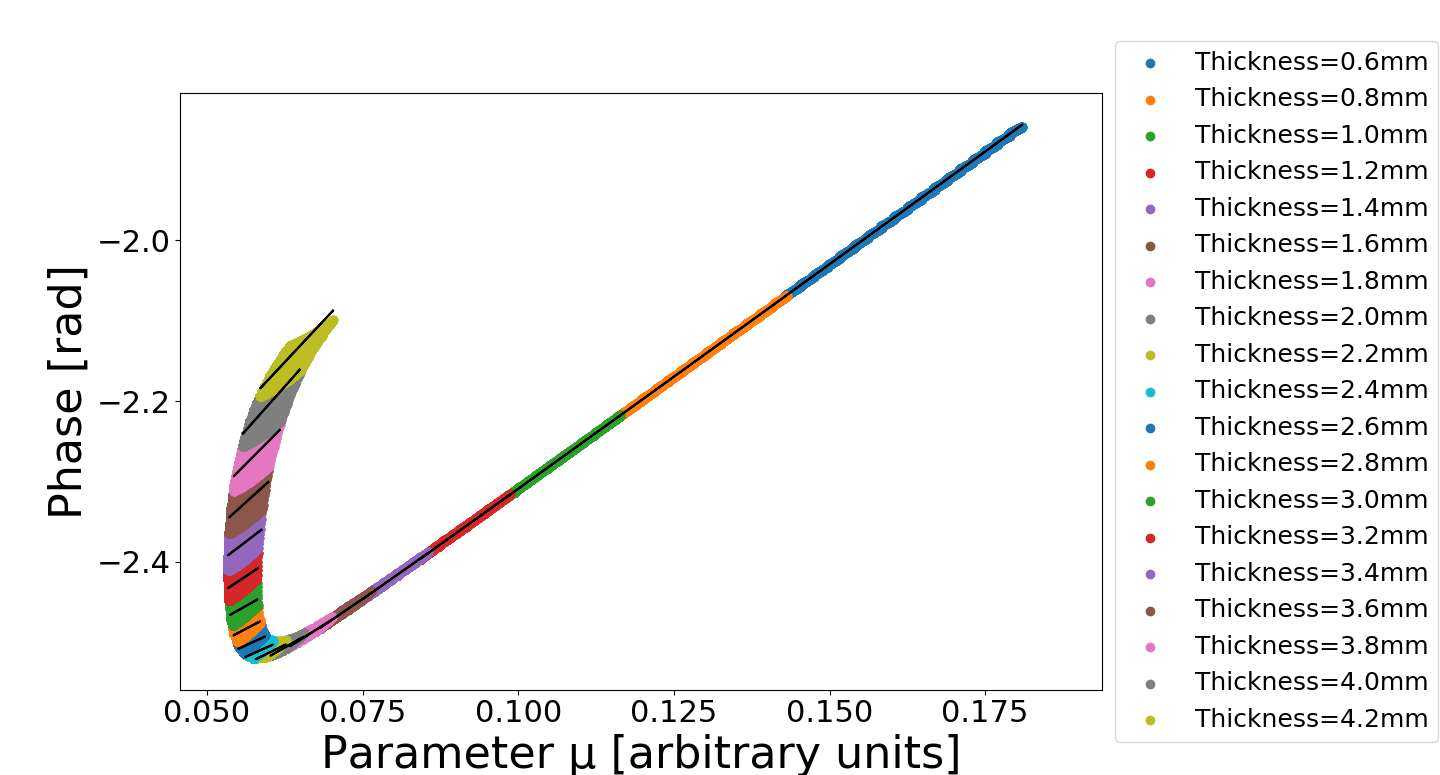}
\caption{Graph of Phase [rad] against Parameter $\mu$ [arbitrary unit] at different thickness}
\label{fig:PhaseMu}
\end{figure}

By plotting the graphs of phase against Parameter $\mu$ at each value of thickness ranging from 0.4~mm to 4.4~mm (Figure \ref{fig:PhaseMu}), we found there is a linear relationship between phase and parameter $\mu$ for each thickness (Figure \ref{fig:PhaseMu}, black lines) and the gradients of the lines are about the same. Both experiment data and simulation data show similar trends.

The linear relationship could be expressed as follows:

\begin{equation}
\phi = m \cdot \mu + c(\delta)
\label{eq:main}
\end{equation}

where $\phi$ is the phase, $m$ is the common gradient, $\mu$ is Parameter $\mu$, $\delta$ is the thickness, $c(\delta)$ is a function of $\delta$ that represents the y-intercept of each straight line. Therefore, the next step is to find the value of $m$ and the expression of $c(\delta)$.

\begin{figure}[H]

\begin{subfigure}{0.5\textwidth}
\includegraphics[width = \linewidth,height=50mm]{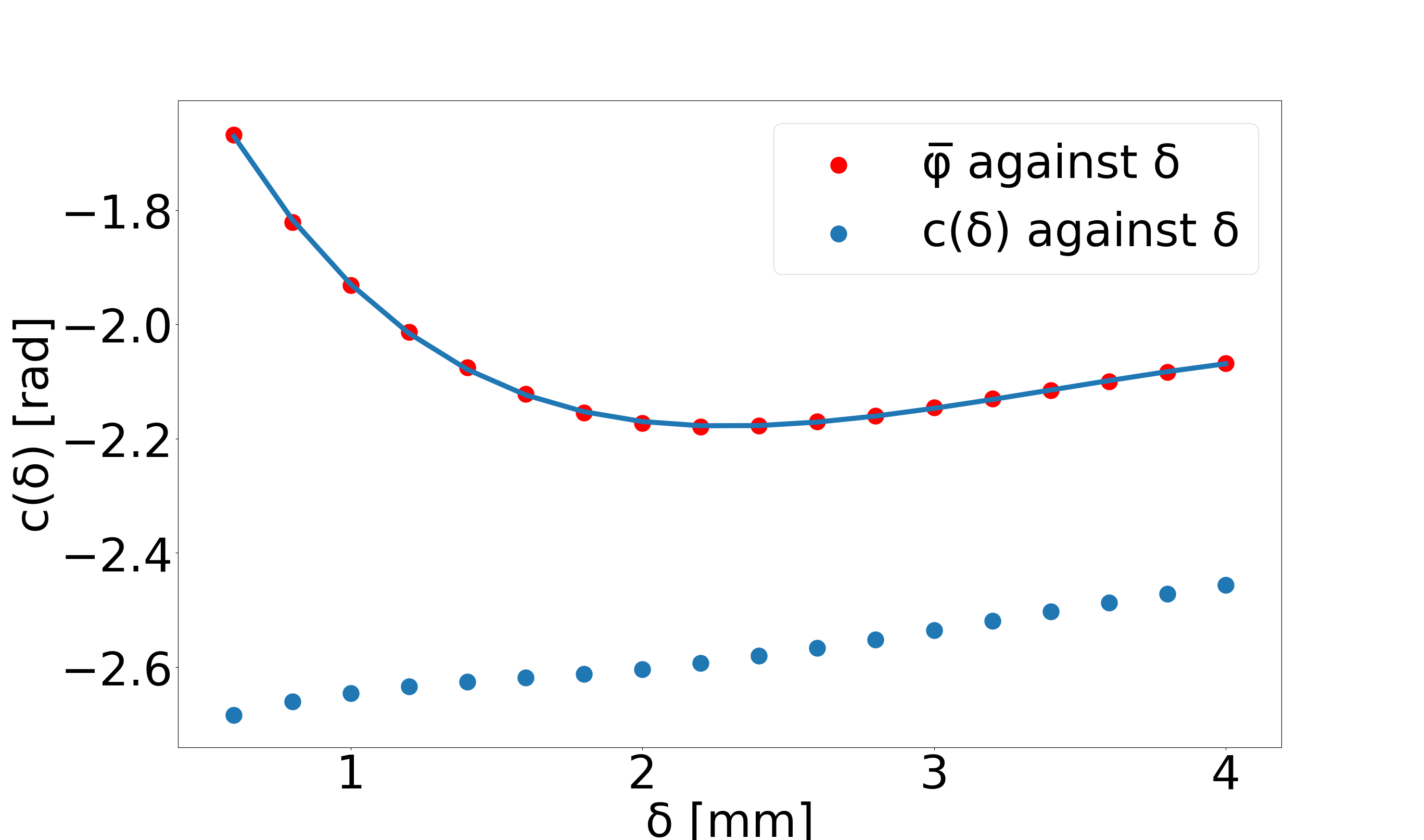} 
\caption{Graphs of $\overline{\phi}$ and $c(\delta)$ [rad] against $\delta$ [mm]}
\label{fig:Polynomial}
\end{subfigure}
\begin{subfigure}{0.5\textwidth}
\includegraphics[width = \linewidth,height=50mm]{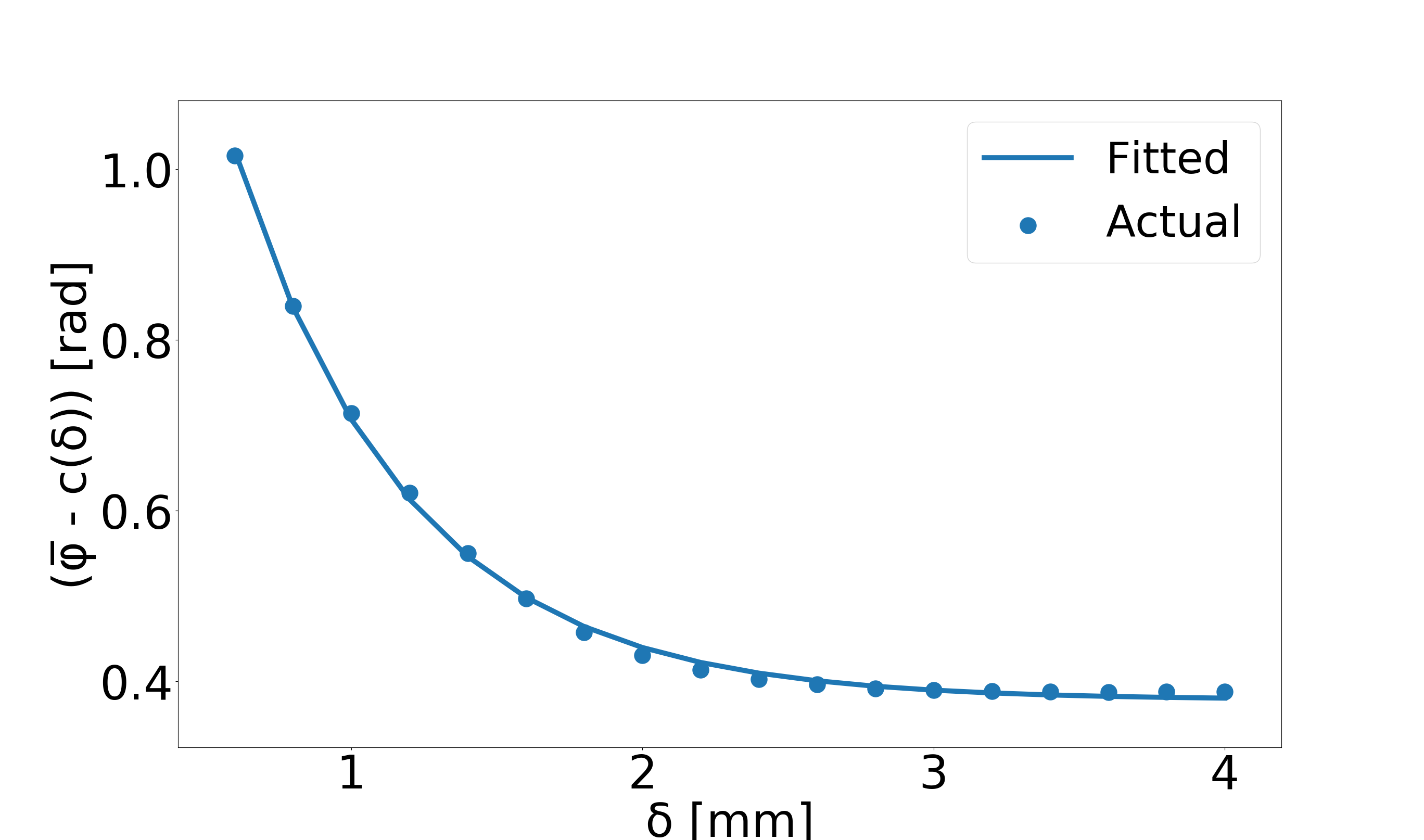}
\caption{Graph of $\overline{\phi} - c(\delta)$ [rad] against $\delta$ [mm]}
\label{fig:Exponential}
\end{subfigure}

\caption{Terms in the equation have an exponential and a polynomial relationship with $\delta$}
\end{figure}

As the lines at different values of thickness have a similar gradient, we took the average of them to be the value of $m$. To find the expression of $c(\delta)$, we plotted $c(\delta)$ against $\delta$ but the trend is not obvious enough to enable the fitting of a known function into it. However, when we plotted $\overline{\phi}$ (i.e. the mean phase value) against thickness, which can be fitted into a five-degree polynomial function, $f(\delta)$, on the same figure (Figure~\ref{fig:Polynomial}), we noticed that the disparity between $\overline{\phi}$ and $c(\delta)$ gradually decreases and finally becomes constant at a certain value. Hence, we plotted $\overline{\phi} - c(\delta)$ against $\delta$ (Figure~\ref{fig:Exponential}, dot) and found that the curve (Figure~\ref{fig:Exponential}, line) can be fitted to a exponential function, $g(x)$. 

Hence, Equation~\ref{eq:main} can now be written as:

\begin{equation}
\phi = m \cdot \mu + f(\delta) - g(\delta)
\label{eq:mainnew}
\end{equation}

As the value of $\phi$ and $\mu$ can both be obtained by processing the raw thermal signal, the value of thickness can be calculated by using Equation~\ref{eq:mainnew} inversely.

\subsection{Technique \text{II}: Building a statistical database}

Quantitative method can also be used for the estimation of thickness by building a sparse database from the measurement and use the database to map thickness. Each data point in the database contains the relevant parameters and if the parameters of a tested data point (except thickness, as it is still unknown) match one sparse data point, we predict the thickness of the tested data point according to the thickness of the sparse data point. 
Before generating the sparse database, we make an evaluation on the set of parameters to examine which parameters are relatively more critical by comparing the multi-dimensional distance between them. We found that Lock-In phase, Lock-In amplitude and 3 other parameters, $\mu_1$, $\mu_2$ and $\mu_3$ are more correlated with thickness. Therefore, our database contains 6 dimensions - thickness, Lock-In phase, Lock-In amplitude, $\mu_1$, $\mu_2$ and $\mu_3$.

When building the database, we gather the data points whose difference is within the set resolution for 6 dimensions into one group and take the average of all the parameters of these data points as a new data point. Note that every data point used has an equal weight. Detailed process of building the database is attached in Appendix~\ref{sec:Technique2}.

All data points in the database are retrieved from the database when estimating thickness of a pixel. 5-dimensional distance is calculated with Equation~\ref{eq:distance} and the data point with the shortest 5-dimensional distance is selected. The corresponding thickness of the selected data point gives the estimated thickness of the pixel. 

\begin{equation}
D = \sqrt{(\phi_t - \phi_b)^2 + (A_t - A_b)^2 + (\mu_{1t} - \mu_{1b})^2 + (\mu_{2t} - \mu_{2b})^2 + (\mu_{3t} - \mu_{3b})^2}
\label{eq:distance}
\end{equation}

In this technique, the choice of resolution is very important. A coarse resolution increases the speed of fitting the geometries, as more data points will be closer than the resolution and hence there will be fewer sparse data points. However, precisely because the resolution is coarse, it will lower the accuracy of the database in mapping thickness. On the other hand, a fine resolution despite being more accurate takes more time to fit the geometries, which lowers the efficiency of the technique. Therefore, we are still looking for an optimal resolution that achieve a balance between accuracy and efficiency. Currently the best resolution set that we have found for (thickness, Lock-In amplitude, Lock-In phase, $\mu_1$, $\mu_2$, $\mu_3$) is (0.2, 50, 0.01, 0.001, 50, 0.01) for experiments and (0.2, 0.1, 0.01, 0.001, 0.1, 0.01) for simulations.

\section{Results and Discussion}

\subsection{Assessment of Technique \text{I}}

\begin{figure}[H]
\begin{subfigure}{0.5\textwidth}
\includegraphics[width = \linewidth,keepaspectratio]{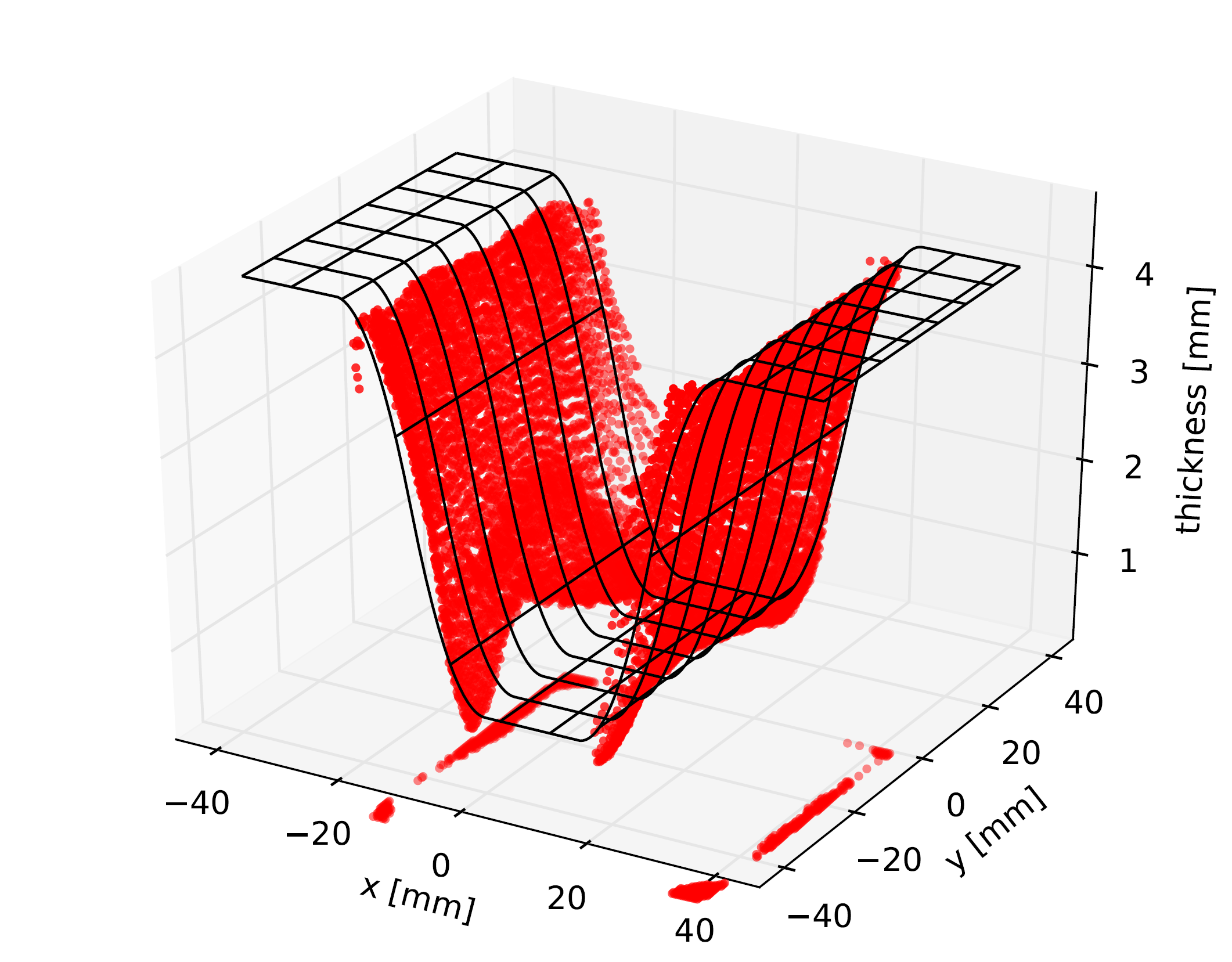} 
\caption{Experiments}
\end{subfigure}
\begin{subfigure}{0.5\textwidth}
\includegraphics[width = \linewidth,keepaspectratio]{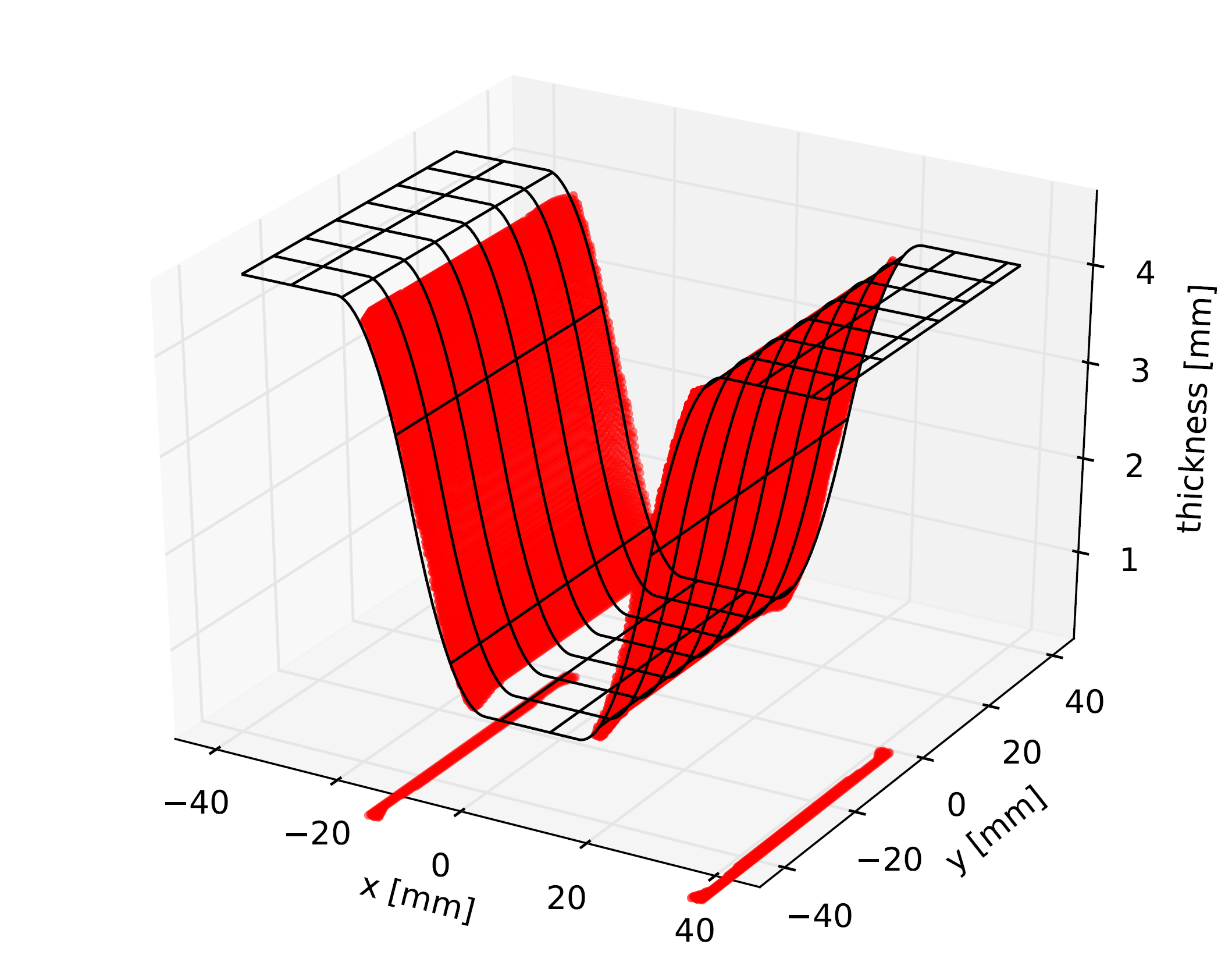}
\caption{Simulations}
\end{subfigure}
\caption{Estimated geometries using Technique \text{I}}
\label{fig:t1}
\end{figure}

To assess Technique \text{I}, we use each sample to build a function and then test the function using all the other samples by calculating the RMSD of the estimated thickness from the actual thickness. The tables of RMSD values for both experiments and simulations are is attached in Appendix~\ref{sec:RMSD}. From the table, we can conclude that Technique \text{I} is not that successful in thickness estimation for simulations. This is evident from the fact that 64 out of 196 values of RMSD is higher than 1 and the highest RMSD value reaches 2.375. Since RMSD has the same unit as the tested variable, which is thickness in this case, the result shown is beyond the range of acceptance for most samples with maximum thickness of 4.4~mm. This is because Technique \text{I} is not applicable to fit the datapoints on the flat surface as in this region many phase values are mapped to by only one thickness value, hence no function can be built. By plotting the estimated geometries of a sample for experiments and simulations are shown below (Figure~\ref{fig:t1}) where the flat region is excluded, we are able to visualise that Technique \text{II} works well for slopes.

\subsection{Assessment of Technique \text{II}}

\begin{figure}[H]
\begin{subfigure}{0.5\textwidth}
\includegraphics[width = \linewidth,keepaspectratio]{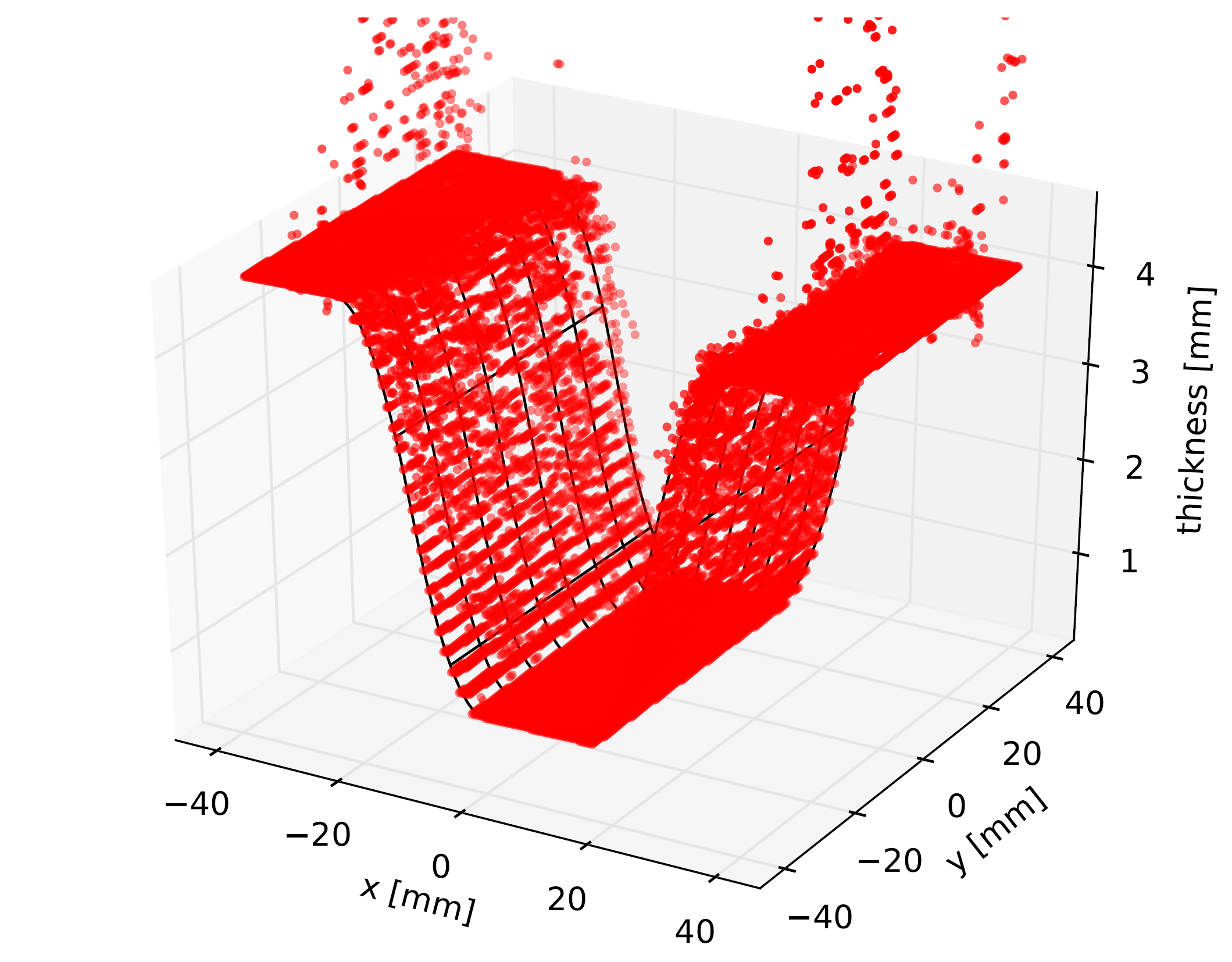} 
\caption{Experiments}
\end{subfigure}
\begin{subfigure}{0.5\textwidth}
\includegraphics[width = \linewidth,keepaspectratio]{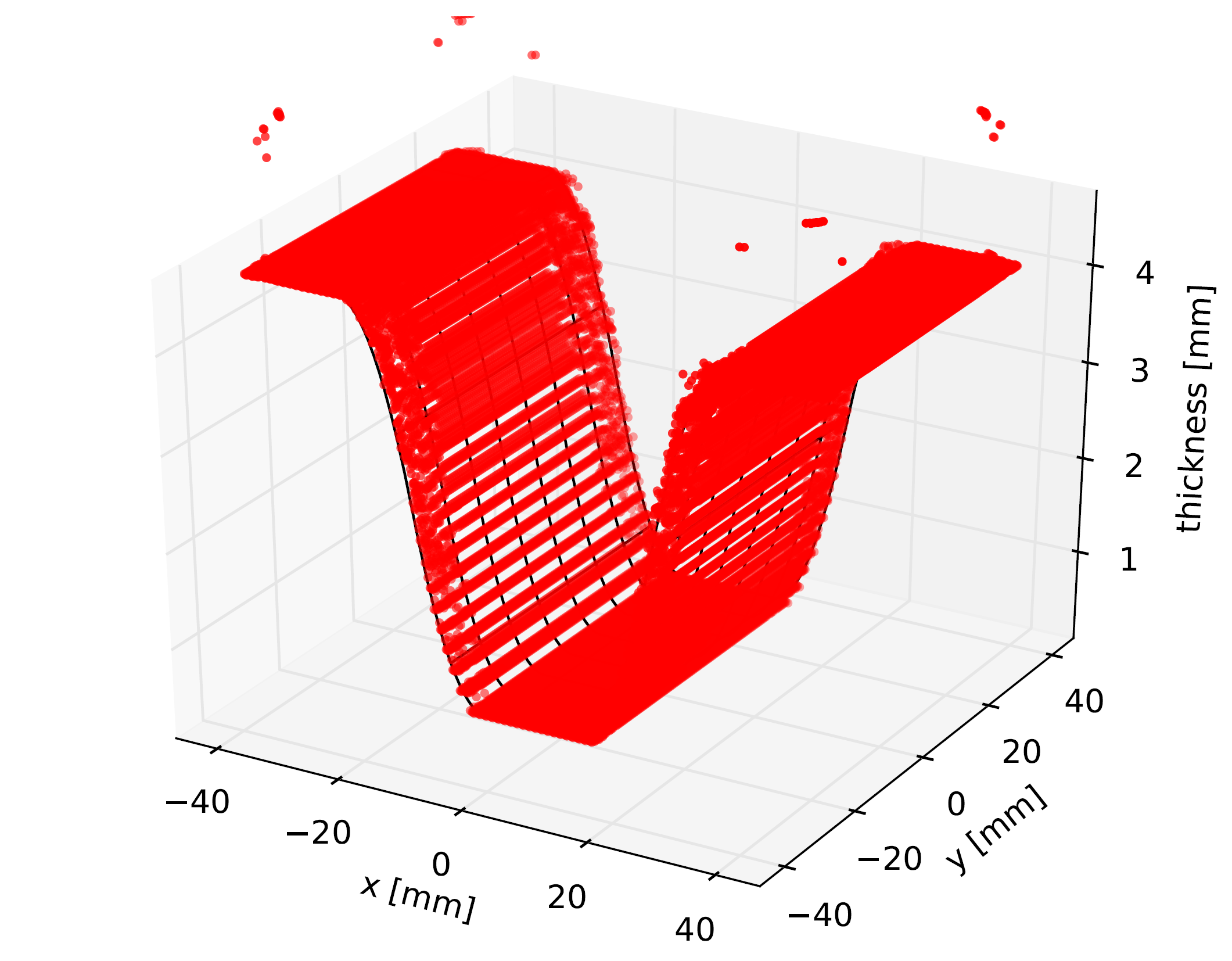}
\caption{Simulations}
\end{subfigure}
\caption{Estimated geometries using Technique \text{II}}
\label{fig:t2}
\end{figure}

To assess Technique \text{II}, we use the 14 samples as tested object and use the database that we have built to estimate the thickness of these tested samples at each pixel. Then we calculate the successful rate of thickness estimation with a set tolerance to have a clear view of the performance of Technique \text{II}. The tables of successful rate for thickness estimation using Technique \text{II} for both experiments and simulations data are attached in Appendix~\ref{sec:SR1}. We can observe from the result that the successful rate of Technique \text{II} reaches around 98\% for simulations and around 83\% for experiments with 0.2~mm. Therefore, Technique \text{II} is successful in thickness estimation. The estimated geometries of a sample for experiments and simulations using Technique \text{II} are shown below (Figure~\ref{fig:t2}).

\subsection{Improvement on Technique \text{II}}

To improve Technique \text{II}, Principal Component Analysis (PCA) is used to reduce the number of dimensions. By analysing the nature of each dimension and removing dimensions providing similar information, the size of the database is reduced while losing the least amount of information \cite{Jolliffer2011IES}. Firstly, data in the raw dataset are normalised, so that for every one of the six dimensions, the mean value of the data in the dimension equals 0 and the standard deviation of the data in the dimension equals 1. This is to make PCA independent of scaling. Secondly, the normalised covariance between every two dimensions is calculated with Equation~\ref{eq:cov}:

\begin{equation}
cov(X,Y) = \frac{\sum_{i = 1}^{n}(X_i - \bar{X})(Y_i - \bar{Y})}{(n - 1)}
\label{eq:cov}
\end{equation}

Normalised covariance measures how much two dimensions vary from the mean value with respect to each other \cite{Smith2002}. The magnitude of the normalised covariance shows the strength of the relationship. If the absolute value normalised covariance is close to one, it means that the two dimensions have strong linear relationship and thus provide similar information. The normalised covariance between every two of the six dimensions for 14 samples is attached in Appendix~\ref{sec:COV}. By observing from the result, we found that for simulation both $\mu_1$ and $\mu_2$ can be removed from the database to reduce the size of it, while for experiments there is only strong linear relationship between amplitude and $\mu_2$, possibly due to existence of thermographic data noise. Hence, for the practical work in the real world, we choose to remove only one dimension, $\mu_2$, from the database. 

The successful rate of thickness estimation is calculated with the modified database. The tables of successful rate for thickness estimation using improved Technique \text{II} for both experiments and simulations data are attached in Appendix~\ref{sec:SR2}. We can observe from the result that Technique \text{II} still performs well, with around 97\% successful rate for simulations and 78\% successful rate for experiments with 0.2~mm tolerance. Therefore, the modification is successful as it reduces the size of the database and thus makes the retrieval of data points from the database faster, while at the same time, the successful rate is not affected much. 

\subsection{Using stochastic gradient descent to identify geometries in real time}

During actual application of the techniques, we do not know how many frames we actually measure to get an accurate result. Hence, we investigated the use of stochastic gradient descent (SGD) in order to determine the time when sufficient data have been collected. 

Gradient descent is a popular iterative optimisation method \cite{Sebastian2011}\cite{Alan2004}\cite{Karl2009} to find the minimum of loss functions. We started from an initial guess of the parameters, and then iteratively moved in the direction of the negative gradient at the current point to reach the minimum value. Out of all kinds of gradient descent, SGD is one that run through every single set of data in the dataset \cite{Stefan2009}\cite{Gardner1984}. Based on this characteristic, whenever we get a new frame, we can use the new set of data to update the parameters and finally reach the set of parameters that lead to a minimum value of the loss function. 

However, during the actual test of this method, we found that one significant drawback of SGD is that the value always converges to the local minimum points instead of the desired global minimum point. Hence, it is important to set the initial guess close enough to the global minimum point (i.e. in this case, the actual value of the parameters). Even though it is possible to use SGD to generate the thickness in real time, the time required to benchmark such a technique is significant. It will be interesting to derive a reliable method to obtain the geometry of samples in real time in future work.

\section{Conclusion}
This project has shown the ability of estimating the thickness of the objects tested under LIT. From the results of accuracy assessment shown by the two methods, we can conclude that Technique \text{II}, which shows optimistic result for both simulations and experiments, performs better in thickness estimation. Technique \text{I} despite performing well for slopes are not able to estimate the thickness of flat regions. Hence, further studies could consider adding in more parameters other than phase and thickness to build equations even for flat surfaces. For Technique \text{II}, the resolution set for each dimension can be adjusted to find an optimal combination of resolution so as to generate a database with the most suitable size. Moreover, we can take more aspects into consideration when building the database. More meaningful dimensions can be added to the database to make Technique \text{II} more mature.

\bibliographystyle{ieeetr}
\break
\bibliography{report}
\break

\renewcommand{\thesection}{Appendix\arabic{section}}
\section*{Appendix}
\setcounter{figure}{0}    
\section{Details of samples used in experiments and simulations}
\label{sec:Details of samples}
\begin{figure}[H]
\centering
\includegraphics[width = 120mm, keepaspectratio]{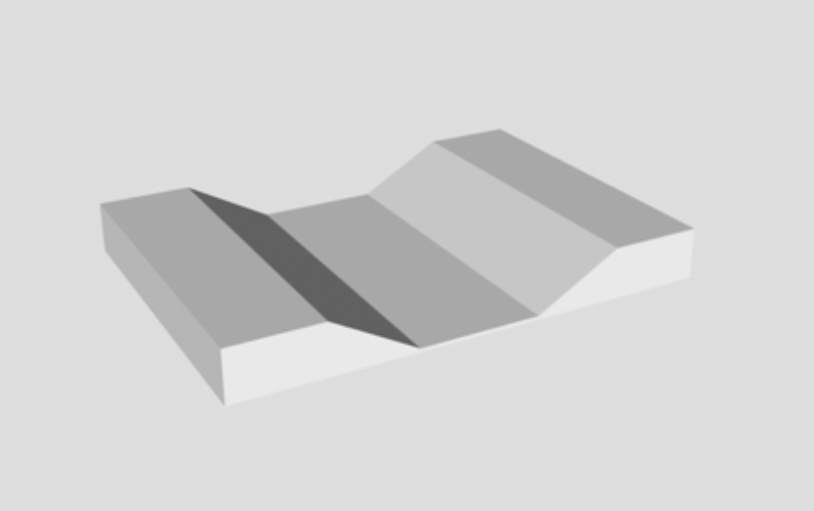}
\caption{Designed model of one of the samples}
\end{figure}

\begin{figure}[H]
\begin{subfigure}{0.5\textwidth}
\includegraphics[width = \linewidth,keepaspectratio]{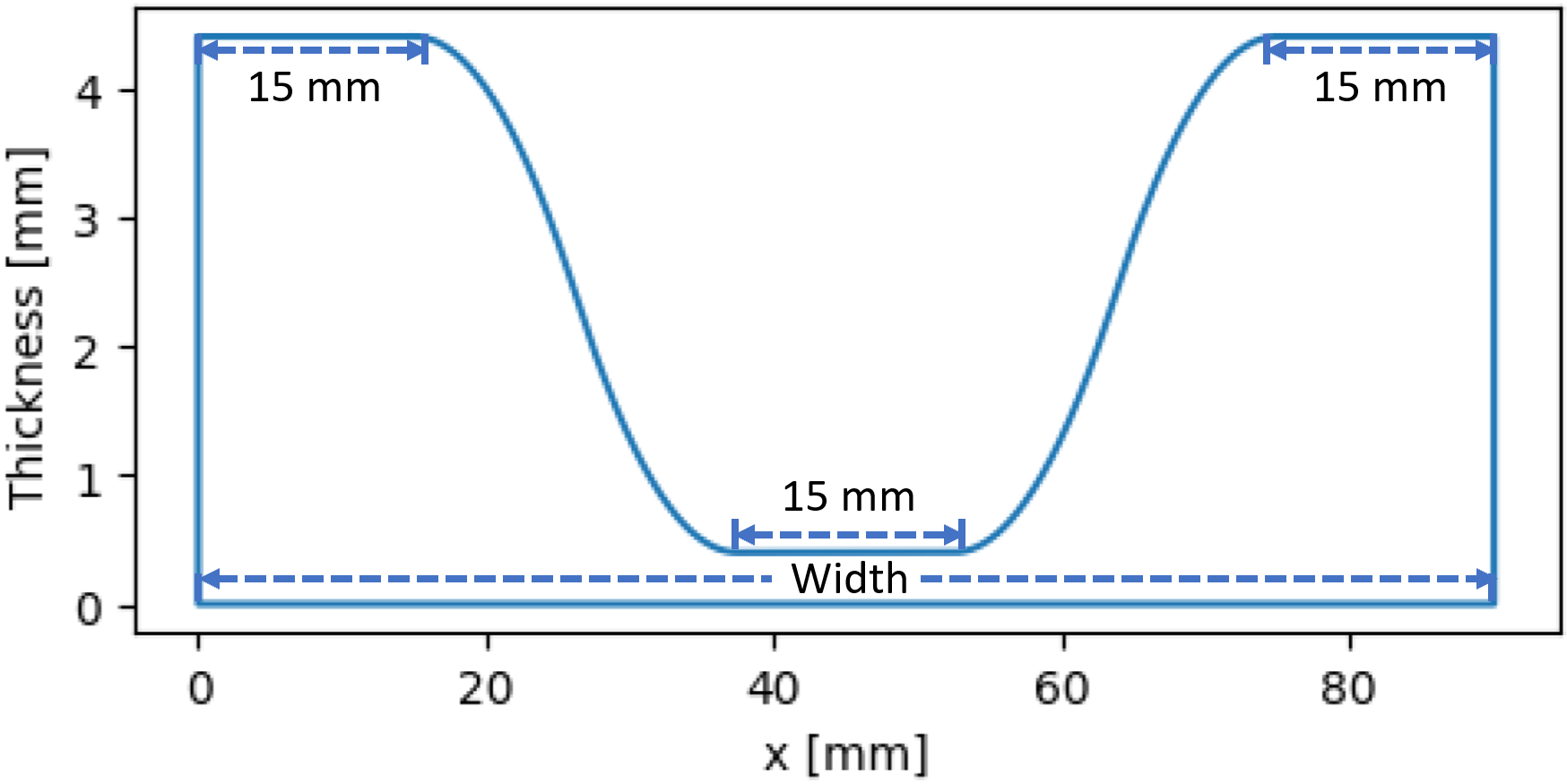} 
\caption{Cross section of a quadratic sample}
\end{subfigure}
\begin{subfigure}{0.5\textwidth}
\includegraphics[width = \linewidth,keepaspectratio]{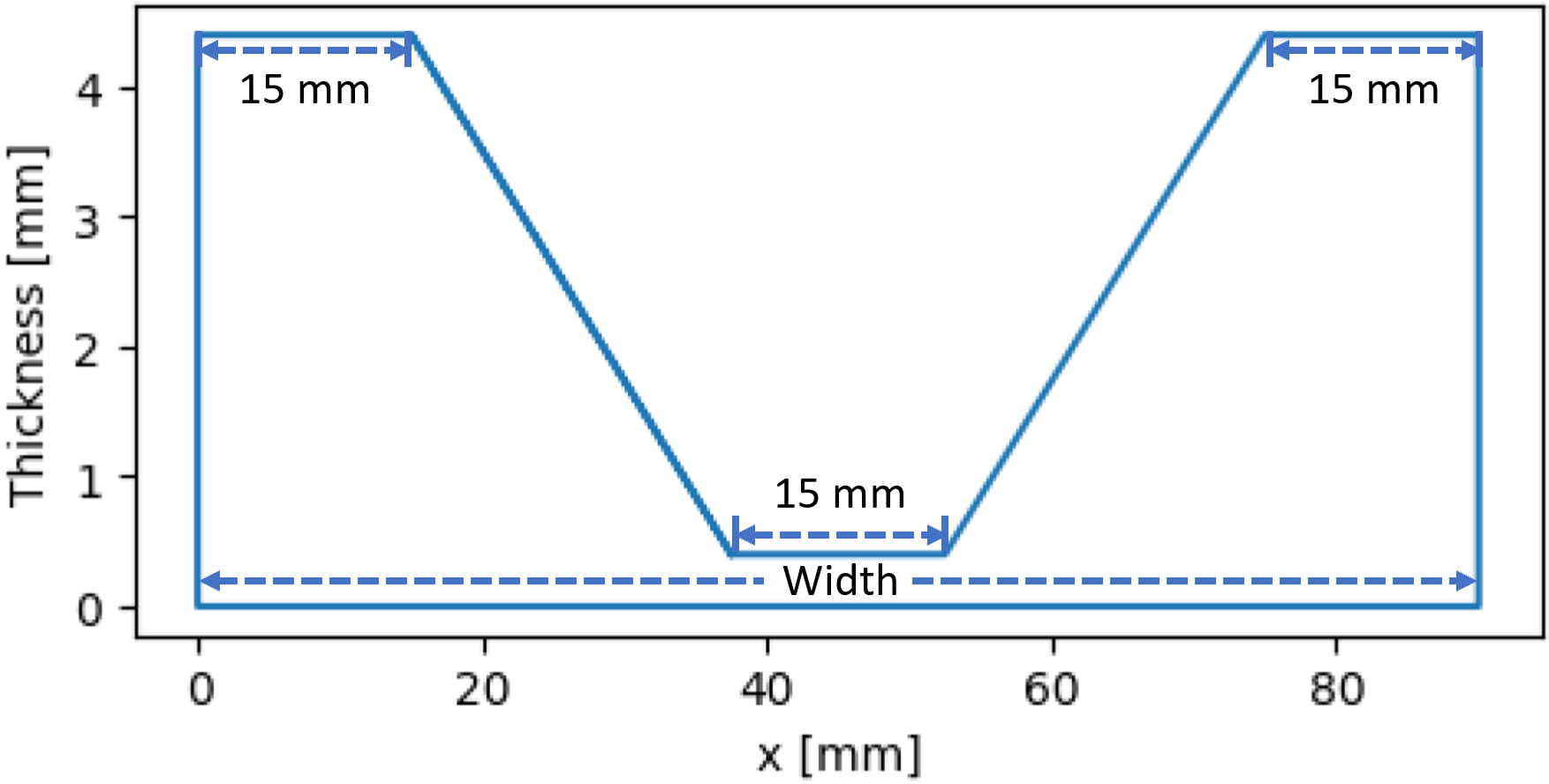}
\caption{Cross section of a linear sample}
\end{subfigure}
\caption{Design of samples} 
\end{figure}

\begin{table}[h]
\begin{center}
\resizebox{\textwidth}{!}{
\begin{tabular}{ | p{1.1cm} |  p{3.2cm} |  p{3.2cm} |  p{3.2cm} |  p{3.2cm} | p{3.2cm} |  p{3.2cm} |}
\hline
Sample & Width (mm) & Max. thickness (mm) & Min. thickness [mm]& Length [mm] &  Max. gradient & Gradient type \\  \hline
1 & 90 & 4.4 & 0.4 & 56 & 0.354 & Quadratic \\
2 & 105 & 4.4 & 0.4 & 56 & 0.267 & Quadratic \\
3 & 75 & 4.4 & 0.4 & 56 & 0.533 & Quadratic \\
4 & 80 & 4.4 & 0.4 & 56 & 0.455 & Quadratic \\
5 & 85 & 4.4 & 0.4 & 56  & 0.400 & Quadratic \\
\hline
6 & 90 & 4.4 & 0.4 & 56 & 0.178 & Linear \\
7 & 105 & 4.4 & 0.4 & 56 & 0.133 & Linear \\
8 & 75 & 4.4 & 0.4 & 56 & 0.267 & Linear \\
9 & 80  & 4.4 & 0.4 & 56 & 0.229 & Linear \\
10 & 85 & 4.4 & 0.4 & 56 & 0.200 & Linear \\ 
11 & 60 & 4.4 & 0.4 & 56 &  0.533 & Linear \\
12 & 65 & 4.4 & 0.4 & 56 & 0.400 & Linear \\
13 & 70 & 4.4 & 0.4 & 56 & 0.320 & Linear \\
14 & 75 & 8.4 & 0.4 & 56 & 0.533 & Linear \\ 
\hline
\end{tabular}
} 
\caption{Details of the 14 samples designed}
\label{tab:sampledetail}
\end{center}
\end{table}

\break

\section{Experiment setup}
\label{sec:Setup}
\begin{figure}[H]
\centering
\includegraphics[width = 120mm, keepaspectratio]{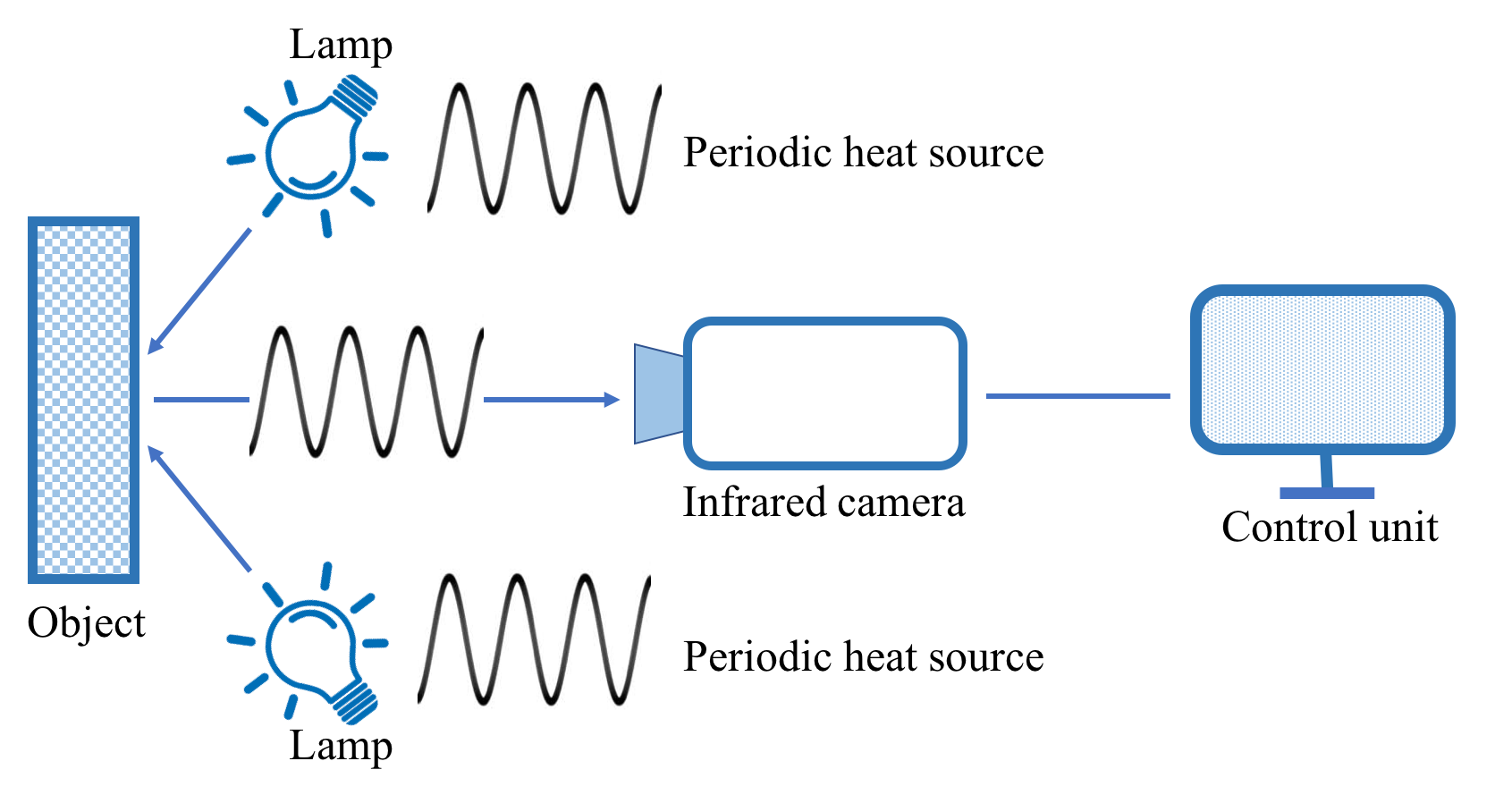}
\caption{A schematic configuration of the LIT experiment setup}
\end{figure}

\break

\section{Equations used for Lock-In calculation}
\label{sec:Lock-In calculation}
\begin{equation}
K_0(t_n)=sin(2\pi ft_n)
\label{Eq:Number1}
\end{equation}
\begin{equation}
K_{-90}(t_n)=-cos(2\pi ft_n)
\label{Eq:Number2}
\end{equation}
\begin{equation}
\label{Eq:Number3}
\begin{split}
S_0 (r,c) & =\sum_{n=1}^{N}(I_n (r,c) K_0 (t_n) \\ 
& =\sum_{n=1}^{N}(I_n (r,c))sin(2\pi ft_n) 
\end{split}
\end{equation}
\begin{equation}
\label{Eq:Number4}
\begin{split}
S_{-90} (r,c) & =\sum_{n=1}^{N}(I_n (r,c) K_{-90} (t_n) \\ 
& =-\sum_{n=1}^{N}(I_n (r,c))cos(2\pi ft_n) 
\end{split}
\end{equation}
\begin{equation}
A(r,c)=\frac{\Delta (t)}{T}\sqrt{(S_0(r,c))^2 +(S_{-90}(r,c))^2 }
\label{Eq:Number5}
\end{equation}
\begin{equation}
\phi(r,c)=\taninv{(\frac{-S_{-90}(r,c)}{S_0(r,c)})}
\label{Eq:Number6}
\end{equation}

\break

\section{Thermal signal plot}
\label{sec:signal}
\begin{figure}[H]
\centering
\includegraphics[width = 100mm,height=60mm]{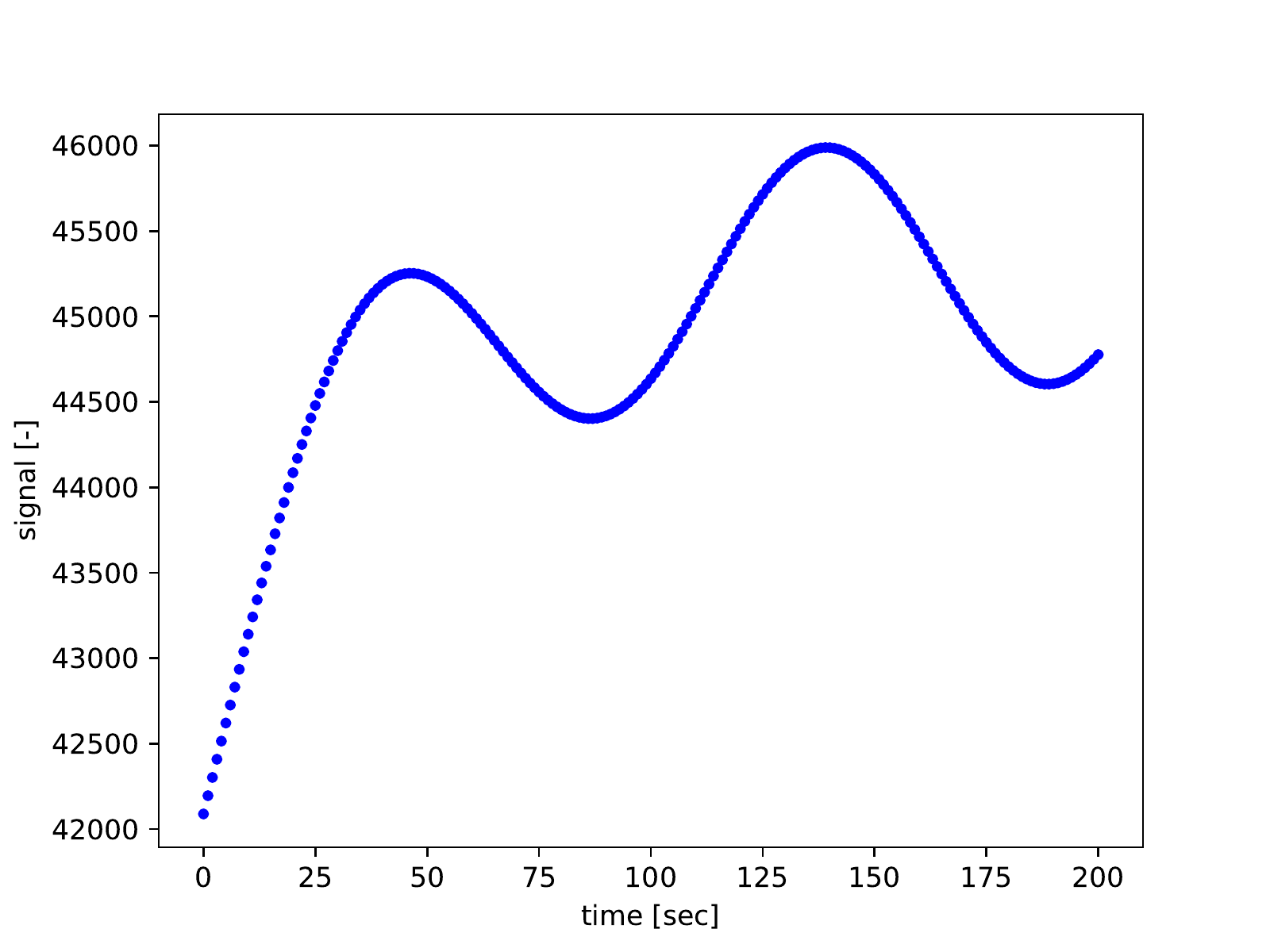}
\caption{A typical thermal signal plot for a single pixel over the measurement period of 200 s}
\label{fig:sigtime}
\end{figure}

\break

\section{Detailed steps of Technique \text{II}}
\label{sec:Technique2}
\begin{figure}[H]
\centering
\includegraphics[height = 200mm, keepaspectratio]{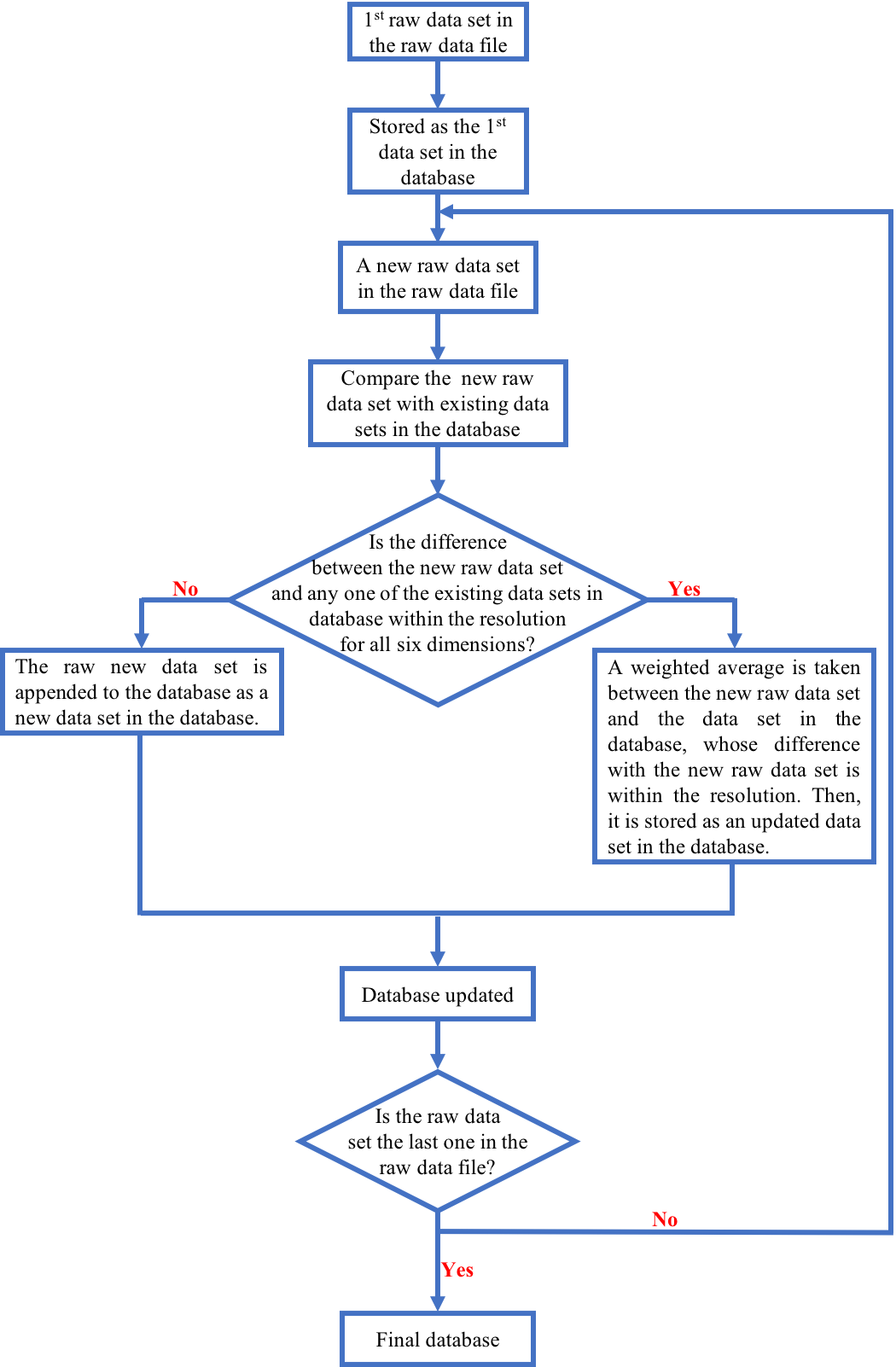}
\caption{Steps to build the database}
\end{figure}

\break

\section{Results of the assessment of Technique \text{I} }
\label{sec:RMSD}
\begin{table}[H]
\centering
\resizebox{\textwidth}{!}{
\begin{tabular}{|c|c|c|c|c|c|c|c|c|c|c|c|c|c|c|c|c|} \hline
 & 1 & 2 & 3 & 4 & 5 & 6 & 7 & 8 & 9 & 10 & 11 & 12 & 13 & 14 \\ \hline
1 & 0.27 & 0.58 & 0.38 & 0.23 & 0.15 & 0.79 & 0.74 & 0.78 & 0.82 & 0.82 & 0.55 & 0.37 & 0.26 & 0.54      \\ \hline
2 & 0.11 & 0.07 & 0.37 & 0.29 & 0.22 & 0.21 & 0.19 & 0.18 & 0.20 & 0.23 & 0.50 & 0.31 & 0.19 & 0.50 \\ \hline
3 & 0.92 & \HighlightBad{1.15} & 0.42 & 0.81 & 0.94 & \HighlightBad{1.04} & \HighlightBad{1.02} & \HighlightBad{1.09} & \HighlightBad{1.07} & \HighlightBad{1.10} & 0.63 & 0.95 & \HighlightBad{1.07} & 0.67   \\ \hline   
4 & 0.76 & 0.78 & 0.26 & 0.46 & 0.73 & 0.94 & 0.91 & 0.98 & 0.98 & 0.95 & 0.58 & 0.33 & 0.93 & 0.60  \\ \hline
5 & 0.61 & 0.68 & 0.31 & 0.20 & 0.47 & 0.86 & 0.82 & 0.89 & 0.90 & 0.87 & 0.56 & 0.37 & 0.75 & 0.56  \\ \hline
6 & 0.98 & \HighlightBad{1.13} & 0.37 & 0.42 & 0.79 & 0.35 & 0.52 & 0.15 & 0.10 & 0.11 & 0.53 & 0.36 & 0.23 & 0.525  \\ \hline
7 & 0.91 & 0.99 & 0.39 & 0.41 & 0.76 & 0.09 & 0.44 & 0.20 & 0.15 & 0.14 & 0.54 & 0.39 & 0.28 & 0.55  \\ \hline
8 & \HighlightBad{1.32} & \HighlightBad{1.52} & 0.36 & 0.97 & \HighlightBad{1.33} & \HighlightBad{1.24} & \HighlightBad{1.34} & 0.56 & \HighlightBad{1.00} & \HighlightBad{1.08} & 0.85 & 0.39 & 0.19 & 0.88  \\ \hline
9 & \HighlightBad{1.23} & \HighlightBad{1.41} & 0.45 & 0.79 & \HighlightBad{1.02} & 0.97 & \HighlightBad{1.07} & 0.37 & 0.61 & 0.75 & 0.76 & 0.63 & 0.53 & 0.75  \\ \hline
10 & \HighlightBad{1.10} & \HighlightBad{1.30} & 0.39 & 0.64 & 0.80 & 0.65 & 0.87 & 0.16 & 0.29 & 0.64 & 0.57 & 0.41 & 0.27 & 0.60  \\ \hline
11 & \HighlightBad{2.28} & \HighlightBad{2.38} & \HighlightBad{2.02} & \HighlightBad{2.15} & \HighlightBad{2.19} & \HighlightBad{2.21} & \HighlightBad{2.22} & \HighlightBad{2.11} & \HighlightBad{2.18} & \HighlightBad{2.24} & 0.24 & \HighlightBad{1.59} & \HighlightBad{1.99} & 0.48  \\ \hline
12 & \HighlightBad{1.19} & \HighlightBad{2.06} & \HighlightBad{1.23} & \HighlightBad{1.63} & \HighlightBad{1.83} & \HighlightBad{1.84} & \HighlightBad{1.86} & \HighlightBad{1.68} & \HighlightBad{1.77} & \HighlightBad{1.79} & 0.45 & 0.09 & \HighlightBad{1.43} & 0.47 \\ \hline
13 & \HighlightBad{1.52} & \HighlightBad{1.79} & 0.82 & \HighlightBad{1.20} & \HighlightBad{1.32} & \HighlightBad{1.53} & \HighlightBad{1.63} & \HighlightBad{1.25} & \HighlightBad{1.39} & \HighlightBad{1.43} & 0.54 & 0.25 & 0.34 & 0.58  \\ \hline
14 & \HighlightBad{2.02} & \HighlightBad{2.14} & \HighlightBad{1.68} & \HighlightBad{1.85} & \HighlightBad{1.96} & \HighlightBad{1.92} & \HighlightBad{1.94} & \HighlightBad{1.78} & \HighlightBad{1.85} & \HighlightBad{1.88} & 0.07 & \HighlightBad{1.02} & \HighlightBad{1.60} & 0.06 \\ \hline
\end{tabular}
}
\caption{RMSD value (2d.p.) of estimated thickness from actual thickness (Simulations). Rows correspond to training data and columns correspond to the test data.}
\end{table}

\begin{table}[H]
\centering
\resizebox{\textwidth}{!}{
\begin{tabular}{|c|c|c|c|c|c|c|c|c|c|c|c|c|c|c|c|c|} \hline
 & 1 & 2 & 3 & 4 & 5 & 6 & 7 & 8 & 9 & 10 & 11 & 12 & 13 & 14 \\ \hline
1 & 0.29 & 0.26	& 0.22 & 0.16 & 0.16 & 0.15 & 0.13 & 0.37 & 0.30 & 0.17 & 0.45 & 0.46 & 0.25 & 0.32 \\ \hline
2 & 0.87	 & \HighlightBad{1.06} & 0.80 & 0.92 & 0.89 & 0.95 & 0.99 & 0.98 & 0.93 & 0.91 & 0.81 & 0.89 & 0.68 & 0.85 \\ \hline
3 & 0.36 & 0.28 & 0.34 & 0.12 & 0.27 & 0.21 & 0.22 & 0.18 & 0.20 & 0.22 & 0.15 & 0.16 & 0.28 & 0.12 \\ \hline
4 & 0.77	 & 0.39 & 0.79 & 0.31 & 0.60 & 0.18 & 0.16 & 0.33 & 0.17 & 0.26 & 0.19 & 0.13 & 0.34 & 0.19 \\ \hline
5 & 0.62 & 0.39 & 0.64 & 0.18 & 0.46 & 0.16 & 0.16 & 0.45 & 0.34 & 0.19 & 0.50 & 0.49 & 0.31 & 0.35 \\ \hline
6 & 0.36 & 0.30 & 0.36 & 0.17 & 0.20 & 0.16 & 0.17 & 0.32 & 0.18 & 0.15 & 0.36 & 0.31 & 0.21 & 0.26 \\ \hline
7 & 0.41 & 0.38 & 0.36 & 0.28 & 0.25 & 0.26 & 0.22 & 0.41 & 0.39 & 0.25 & 0.48 & 0.50 & 0.34 & 0.39 \\ \hline
8 & 0.45 & 0.32 & 0.48 & 0.15 & 0.34 & 0.18 & 0.17 & 0.14 & 0.19 & 0.17 & 0.22 & 0.12 & 0.23 & 0.21 \\ \hline
9 & 0.58 & 0.35 & 0.61 & 0.25 & 0.46 & 0.13 & 0.14 & 0.21 & 0.13 & 0.21 & 0.33 & 0.14 & 0.28 & 0.28 \\ \hline
10 & 0.50 & 0.55 & 0.40 & 0.45 & 0.42 & 0.48 & 0.42 & 0.52 & 0.55 & 0.41 & 0.56 & 0.62 & 0.46 & 0.50 \\ \hline
11 & \HighlightBad{1.11} & 0.93 & \HighlightBad{1.20} & \HighlightBad{1.03} & \HighlightBad{1.08} & 0.70 & 0.78 & 0.68 & 0.50 & 0.76 & 0.13 & 0.41 & 0.71 & 0.62 \\ \hline
12 & \HighlightBad{1.02} & 0.58 & \HighlightBad{1.00} & 0.59 & 0.86 & 0.44 & 0.46 & 0.61 & 0.36 & 0.51 & 0.39 & 0.32 & 0.52 & 0.42 \\ \hline
13 & 0.33 & 0.30 & 0.37 & 0.19 & 0.19 & 0.14 & 0.16 & 0.29 & 0.15 & 0.15 & 0.31 & 0.21 & 0.17 & 0.27 \\ \hline
14 & 0.77 & 0.44 & 0.84 & 0.47 & 0.65 & 0.30 & 0.29 & 0.34 & 0.27 & 0.36 & 0.16 & 0.22 & 0.41 & 0.17 \\ \hline
\end{tabular}
}
\caption{RMSD value (2d.p.) of estimatd thickness from actual thickness (Experiments). Rows correspond to training data and columns correspond to the test data.}
\end{table}
\break

\section{Results of the assessment of Technique \text{II} }
\label{sec:SR1}
\begin{table}[H]
\centering
\resizebox{\textwidth}{!}{
\begin{tabular}{ | p{1.4cm} | p{2cm} | p{2cm} | p{2cm} | p{2cm} | p{2cm} | p{2cm} | p{2cm} | p{2cm} | }
\hline
Sample & 0.2mm tolerance & 0.4mm tolerance & 0.6mm tolerance & 0.8mm tolerance & 1.0mm tolerance & 1.2mm tolerance & 1.4mm tolerance & 1.6mm tolerance \\  \hline
1 & 98.990\% & 99.779\% & 99.922\% & 99.922\% & 99.922\% & 99.930\% & 99.930\% & 100.00\% \\
2 & 98.990\% & 99.795\% & 99.897\% & 99.911\% & 99.911\% & 99.945\% & 99.945\% & 100.00\% \\
3 & 98.450\% & 99.883\% & 99.909\% & 99.909\% & 99.909\% & 99.919\% & 99.919\% & 100.00\% \\
4 & 97.900\% & 99.906\% & 99.964\% & 99.964\% & 99.964\% & 99.964\% & 99.964\% & 100.00\% \\
5 & 98.175\% & 99.865\% & 99.958\%  & 99.962\% & 99.962\% & 99.971\% & 99.971\% & 100.00\% \\
6 & 99.008\% & 99.843\% & 99.892\% & 99.900\% & 99.902\% & 99.936\% & 99.936\% & 100.00\% \\
7 & 98.745\% & 99.761\% & 99.857\% & 99.857\% & 99.887\% & 99.969\% & 99.973\% & 100.00\% \\
8 & 97.996\% & 99.756\% & 99.890\% & 99.900\% & 99.900\% & 99.907\% & 99.907\% & 100.00\% \\
9 & 98.719\%& 99.839\% & 99.910\% & 99.933\% & 99.933\% & 99.955\% & 99.955\% & 100.00\% \\
10 & 99.322\% & 99.874\% & 99.945\% & 99.954\% & 99.954\% & 99.962\% & 99.962\% & 100.00\% \\ 
11 & 92.182\% & 98.686\% & 99.510\% & 99.809\% & 99.809\% & 99.809\% & 99.809\% & 100.00\% \\
12 & 98.175\% & 99.785\% & 99.851\% & 99.857\% & 99.857\% & 99.857\% & 99.857\% & 100.00\% \\
13 & 97.052\% & 99.744\% & 99.882\% & 99.882\% & 99.882\%& 99.893\% & 99.893\% & 100.00\% \\
14 & 97.883\% & 98.699\% & 98.811\% & 98.850\% & 98.859\% & 99.015\% & 99.990\% & 100.00\% \\ 
\hline
\end{tabular}
}
\caption{Successful rate of Technique \text{II} (Simulations)}
\label{tbl:SR}
\end{table}

\begin{table}[H]
\centering
\resizebox{\textwidth}{!}{
\begin{tabular}{ | p{1.4cm} | p{2cm} | p{2cm} | p{2cm} | p{2cm} | p{2cm} | p{2cm} | p{2cm} | p{2cm} | }
\hline
Sample & 0.2mm tolerance & 0.4mm tolerance & 0.6mm tolerance & 0.8mm tolerance & 1.0mm tolerance & 1.2mm tolerance & 1.4mm tolerance & 1.6mm tolerance \\  \hline
1 & 87.350\% & 94.559\% & 96.603\% & 97.674\% & 98.363\% & 98.764\% & 99.195\% & 99.348\% \\
2 & 85.930\% & 94.673\% & 97.082\% & 98.119\% & 98.659\% & 99.025\% & 99.311\% & 99.476\% \\
3 & 86.054\% & 94.729\% & 97.578\% & 98.567\% & 98.999\% & 99.226\% & 99.485\% & 99.574\% \\
4 & 90.541\% & 96.731\% & 98.493\% & 99.047\% & 99.306\% & 99.446\% & 99.531\% & 99.614\% \\
5 & 87.749\% & 95.272\% & 97.492\%  & 98.562\% & 99.046\% & 99.334\% & 99.526\% & 99.628\% \\
6 & 84.827\% & 93.194\% & 97.368\% & 98.835\% & 99.335\% & 99.633\% & 99.761\% & 99.831\% \\
7 & 81.673\% & 90.105\% & .94.659\% & 97.228\% & 98.253\% & 98.703\% & 99.120\% & 99.355\% \\
8 & 80.944\% & 90.703\% & 94.691\% & 96.988\% & 98.489\% & 99.108\% & 99.417\% & 99.578\% \\
9 & 86.376\%& 93.658\% & 96.576\% & 98.263\% & 99.108\% & 99.471\% & 99.663\% & 99.758\% \\
10 & 84.720\% & 93.328\% & 96.452\% & 97.985\% & 98.657\% & 98.971\% & 99.283\% & 99.425\% \\ 
11 & 77.507\% & 87.553\% & 93.209\% & 97.113\% & 98.974\% & 99.605\% & 99.823\% & 99.918\% \\
12 & 88.358\% & 94.012\% & 96.135\% & 97.306\% & 98.076\% & 98.600\% & 98.934\% & 99.171\% \\
13 & 83.337\% & 90.367\% & 94.050\% & 96.185\% & 97.462\%& 97.984\% & 98.476\% & 98.758\% \\
14 & 53.784\% & 65.853\% & 73.861\% & 79.386\% & 83.543\% & 86.136\% & 88.487\% & 90.282\% \\ 
\hline
\end{tabular}
}
\caption{Successful rate of Technique \text{II} (Experiments)}
\label{tbl:SR}
\end{table}
\break

\section{Normalised covariance matrix calculated for the improvement of Technique \text{II} }
\label{sec:COV}
\begin{table}[H]
\centering
\resizebox{\textwidth}{!}{
\begin{tabular}{|*{7}{c|}c|c|*{4}{c|}c|*{2}{c|}}
\hline
Sample & $T\&A$ & $T\&\phi$ & $T\&\mu_1$ & $T\&\mu_2$ & $T\&\mu_3$ & $A\&\phi$ & \HighlightCyan{$A\&\mu_1$} & \HighlightCyan{$A\&\mu_2$} & $A\&\mu_3$ & $\phi\&\mu_1$ & $\phi\&\mu_2$ & $\phi\&\mu_3$ & \HighlightCyan{$\mu_1\&\mu_2$} & $\mu_1\&\mu_3$ & $\mu_2\&\mu_3$ \\ \hline
1 & -0.953 & -0.303 & -0.892 & -0.844 & 0.935 & 0.558 & \HighlightCyan{0.985} & \HighlightCyan{0.964} & -0.847 & 0.680 & 0.740 & -0.189 & \HighlightCyan{0.992} & -0.759 & -0.694 \\
2 & -0.950 & -0.216 & -0.870 & -0.821 & 0.931 &  0.500 & \HighlightCyan{0.978} & \HighlightCyan{0.956} & -0.853 & 0.654 & 0.714 & -0.120 & \HighlightCyan{0.993} & -0.742 & -0.678 \\
3 & -0.955 & -0.444 & -0.917 & -0.875 & 0.945 & 0.646 & \HighlightCyan{0.992} & \HighlightCyan{0.976} & -0.843 & 0.726 & 0.781 & -0.302 & \HighlightCyan{0.991} & -0.786 & -0.725 \\
4 & -0.955 & -0.386 & -0.908 & -0.863 & 0.940 & 0.611 & \HighlightCyan{0.989} & \HighlightCyan{0.972} & -0.843 & 0.706 & 0.764 & -0.256 & \HighlightCyan{0.991} & -0.775 & -0.712 \\
5 & -0.954 & -0.339 & -0.899 & -0.853 & 0.938 & 0.582 & \HighlightCyan{0.987} & \HighlightCyan{0.968} & -0.845 & 0.691 & 0.750 & -0.218 & \HighlightCyan{0.991} & -0.767 & -0.702 \\
6 & -0.938 & -0.145 & -0.839 & -0.778 & 0.904 & 0.459 & \HighlightCyan{0.972} & \HighlightCyan{0.940} & -0.829 & 0.641 & 0.714 & -0.053 & \HighlightCyan{0.990} & -0.696 & -0.613 \\
7 &  -0.932 & -0.068 & -0.813 & -0.749 & 0.904 & 0.407 & \HighlightCyan{0.963} & \HighlightCyan{0.928} & -0.836 & 0.623 & 0.695 & 0.003 & \HighlightCyan{0.991} & -0.678 & -0.594 \\
8 & -0.944 & -0.279 & -0.876 & -0.821 & 0.911 & 0.546 & \HighlightCyan{0.984} & \HighlightCyan{0.957} & -0.822 & 0.675 & 0.747 & -0.151 & \HighlightCyan{0.988} & -0.728 & -0.647 \\
9 & -0.942 & -0.223 & -0.862 & -0.805 & 0.907 & 0.510 & \HighlightCyan{0.979} & \HighlightCyan{0.951} &-0.824 & 0.660 & 0.732 & -0.110 & \HighlightCyan{0.989} & -0.715 & -0.632 \\
10 & -0.941 & -0.182 & -0.851 & -0.792 & 0.905 & 0.482 & \HighlightCyan{0.976} & \HighlightCyan{0.945} & -0.826 & 0.649 & 0.722 & -0.078 & \HighlightCyan{0.989} & -0.705 & -0.622 \\
11 & -0.947 & -0.560 & -0.924 & -0.890 & 0.940  & 0.704 & \HighlightCyan{0.995} & \HighlightCyan{0.981} & -0.836 & 0.762 & 0.816 & -0.372 & \HighlightCyan{0.991} &-0.798 & -0.734 \\
12 & -0.947 & -0.440 & -0.908 & -0.863 & 0.928 & 0.641 & \HighlightCyan{0.992} & \HighlightCyan{0.972} & -0.824 & 0.723 & 0.787 & -0.273 & \HighlightCyan{0.989} & -0.766 & -0.692 \\
13 & -0.946 & -0.348 & -0.892 & -0.840 & 0.917 & 0.588 & \HighlightCyan{0.988} & \HighlightCyan{0.964} & -0.821 & 0.694 & 0.764 & -0.202 & \HighlightCyan{0.988} & -0.744& -0.665 \\
14 & -0.858 & 0.233 & -0.743 & -0.774 & 0.866 & 0.253 & \HighlightCyan{0.978} & \HighlightCyan{0.982} & -0.852 & 0.431 & 0.406 & 0.052 & \HighlightCyan{0.987} & -0.778 & -0.758 \\
\hline
\end{tabular}
}
\caption{Normalised covariance of 6 dimensions for 14 samples (Simulations)}
\end{table}

\begin{table}[H]
\centering
\resizebox{\textwidth}{!}{
\begin{tabular}{|*{8}{c|}c|*{7}{c|}}
\hline
Sample & $T\&A$ & $T\&\phi$ & $T\&\mu_1$ & $T\&\mu_2$ & $T\&\mu_3$ & $A\&\phi$ & $A\&\mu_1$ & \HighlightCyan{$A\&\mu_2$} & $A\&\mu_3$ & $\phi\&\mu_1$ & $\phi\&\mu_2$ & $\phi\&\mu_3$ & $\mu_1\&\mu_2$ & $\mu_1\&\mu_3$ & $\mu_2\&\mu_3$ \\ \hline
1 & -0.858 & -0.757 & -0.829 & -0.824 &0.590 & 0.773 & 0.870 & \HighlightCyan{0.985} & -0.445 & 0.806 & 0.844 & -0.106 & 0.858 & -0.305 & -0.328 \\
2 & -0.886 & -0.767 & -0.782 & -0.840 & 0.566 & 0.831 & 0.850 & \HighlightCyan{0.981} & -0.323 & 0.745 & 0.903 & -0.120 & 0.807 & -0.215 & -0.210 \\
3 & -0.865 & -0.844 & -0.884 & -0.849 & 0.548 & 0.808 & 0.892 & \HighlightCyan{0.990} & -0.473 & 0.879 & 0.863 & -0.127 & 0.896 & -0.325 & -0.367 \\
4 & -0.883 & -0.805 & -0.820 & -0.854 & 0.623 & 0.835 & 0.894 & \HighlightCyan{0.986} & -0.404 & 0.800 & 0.896 & -0.158 & 0.874 & -0.279 & -0.296  \\
5 & -0.852 & -0.781 & -0.534 & -0.814 & 0.714 & 0.786 & 0.534 & \HighlightCyan{0.979} & -0.545 & 0.394 & 0.854 & -0.307 & 0.416 & -0.217 & -0.463 \\
6 & -0.839 & -0.667 & -0.831 & -0.775 & 0.628 & 0.730 & 0.847 & \HighlightCyan{0.978} & -0.449 & 0.826 & 0.823 & -0.024 & 0.847 & -0.319 & -0.287 \\
7 &  -0.848 & -0.626 & -0.665 & -0.766 & 0.692 & 0.731 & 0.705 & \HighlightCyan{0.965} & -0.506 & 0.537 & 0.847 & -0.119 & 0.608 & -0.325 & -0.345  \\
8 & -0.852 & -0.741  & -0.845 & -0.818 & 0.459 & 0.746 & 0.854 & \HighlightCyan{0.982} & -0.396 & 0.833 & 0.833 & 0.032 & 0.864 & -0.222 & -0.254 \\
9 & -0.842 & -0.747 & -0.851 & -0.799 & 0.471 & 0.777 & 0.882 & \HighlightCyan{0.982} & -0.300 & 0.851 & 0.854 & 0.027 & 0.884 & -0.238 & -0.151 \\
10 & -0.836 & -0.696 & -0.777 & -0.783 & 0.547 & 0.770 & 0.840 & \HighlightCyan{0.980} & -0.321 & 0.734 & 0.856 & 0.013 & 0.812 & -0.209 & -0.176 \\
11 & -0.8607 & -0.849 & -0.885 & -0.853 & 0.505 & 0.803 & 0.889 & \HighlightCyan{0.991} & -0.351 & 0.908 & 0.857 & -0.082 & 0.900 & -0.251 & -0.253   \\
12 & -0.848 & -0.810 & -0.616 & -0.814 & 0.537 & 0.823 & 0.561 & \HighlightCyan{0.982} & -0.420 & 0.559 & 0.873 & -0.143 & 0.466 & -0.021 & -0.347 \\
13 & -0.847 & -0.741 & -0.760 & -0.811 & 0.508 & 0.780 & 0.807 & \HighlightCyan{0.983} & -0.331 & 0.708 & 0.852 & 0.004 & 0.763 & -0.168 & -0.217 \\
\hline
\end{tabular}
}
\caption{Normalised covariance of 6 dimensions for 14 samples (Experiments)}
\end{table}
\break

\section{Assessment of Technique \text{II} after improvement}
\label{sec:SR2}
\begin{table}[H]
\centering
\resizebox{\textwidth}{!}{
\begin{tabular}{ | p{1.4cm} | p{2cm} | p{2cm} | p{2cm} | p{2cm} | p{2cm} | p{2cm} | p{2cm} | p{2cm} | }
\hline
Sample & 0.2mm tolerance & 0.4mm tolerance & 0.6mm tolerance & 0.8mm tolerance & 1.0mm tolerance & 1.2mm tolerance & 1.4mm tolerance & 1.6mm tolerance \\  \hline
1 & 98.566\% & 99.763\% & 99.876\% & 99.878\% & 99.878\% & 99.916\% & 99.930\% & 100.00\% \\
2 & 98.531\% & 99.702\% & 99.866\% & 99.880\% & 99.884\% & 99.935\% & 99.945\% & 100.00\% \\
3 & 97.924\% & 99.845\% & 99.854\% & 99.854\% & 99.854\% & 99.890\% & 99.904\% & 100.00\% \\
4 & 98.191\% & 99.879\% & 99.933\% & 99.933\% & 99.933\% & 99.960\% & 99.964\% & 100.00\% \\
5 & 98.289\% & 99.853\% & 99.920\%  & 99.928\% & 99.928\% & 99.966\% & 99.971\% & 100.00\% \\
6 & 98.121\% & 99.753\% & 99.857\% & 99.857\% & 99.863\% & 99.910\% & 99.924\% & 100.00\% \\
7 & 97.414\% & 99.652\% & 99.857\% & 99.860\% & 99.877\% & 99.969\% & 99.973\% & 100.00\% \\
8 & 97.037\% & 99.670\% & 99.849\% & 99.857\% & 99.857\% & 99.885\% & 99.904\% & 100.00\% \\
9 & 98.428\%& 99.767\% & 99.888\% & 99.906\% & 99.910\% & 99.946\% & 99.951\% & 100.00\% \\
10 & 98.904\% & 99.827\% & 99.924\% & 99.933\% & 99.933\% & 99.954\% & 99.958\% & 100.00\% \\ 
11 & 90.868\% & 97.832\% & 99.474\% & 99.797\% & 99.797\% & 99.797\% & 99.803\% & 100.00\% \\
12 & 97.277\% & 99.642\% & 99.829\% & 99.840\% & 99.846\% & 99.851\% & 99.857\% & 100.00\% \\
13 & 96.463\% & 99.713\% & 99.862\% & 99.862\% & 99.862\%& 99.882\% & 99.887\% & 100.00\% \\
14 & 97.802\% & 98.735\% & 98.855\% & 98.876\% & 98.876\% & 99.029\% & 99.995\% & 100.00\% \\ 
\hline
\end{tabular}
}
\caption{Successful rate of Technique \text{II} after improvement (Simulations)}
\label{tbl:SR}
\end{table}

\begin{table}[H]
\centering
\resizebox{\textwidth}{!}{
\begin{tabular}{ | p{1.4cm} | p{2cm} | p{2cm} | p{2cm} | p{2cm} | p{2cm} | p{2cm} | p{2cm} | p{2cm} | }
\hline
Sample & 0.2mm tolerance & 0.4mm tolerance & 0.6mm tolerance & 0.8mm tolerance & 1.0mm tolerance & 1.2mm tolerance & 1.4mm tolerance & 1.6mm tolerance \\  \hline
1 & 85.976\% &	93.693\% &	96.114\% &	97.488\% &	98.109\% &	98.565\% &	99.012\% &	99.189\% \\ 
2 & 84.678\% &	93.775\% &	96.642\% &	97.897\% &	98.659\% &	99.027\% &	99.289\% &	99.482\% \\ 
3 & 84.798\% &	94.059\% &	97.070\% &	98.204\% &	98.733\% &	99.074\% &	99.422\% &	99.581\% \\ 
4 & 89.472\% &	96.504\% &	98.213\% &	98.889\% &	99.222\% &	99.380\% &	99.510\% &	99.604\% \\ 
5 & 85.974\% &	94.395\% &	96.941\% &	98.059\% &	98.658\% &	99.071\% &	99.289\% &	99.455\% \\ 
6 & 83.457\% &	92.510\% &	96.848\% &	98.521\% &	99.286\% &	99.595\% &	99.725\% &	99.825\% \\ 
7 & 79.874\% &	89.415\% &	94.264\% &	96.840\% &	97.897\% &	98.549\% &	99.065\% &	99.300\% \\ 
8 & 78.508\% &	89.793\% &	94.072\% &	96.668\% &	98.128\% &	98.980\% &	99.283\% &	99.505\% \\ 
9 & 84.750\% &	92.759\% &	96.347\% &	98.182\% &	98.929\% &	99.370\% &	99.596\% &	99.714\% \\ 
10 & 83.258\% &	92.360\% &	95.777\% &	97.506\% &	98.435\% &	98.914\% &	99.142\% &	99.338\% \\ 
11 & 74.860\% &	86.300\% &	92.483\% &	96.477\% &	98.597\% &	99.446\% &	99.737\% &	99.891\% \\ 
12 & 86.384\% &	92.617\% &	95.133\% &	96.587\% &	97.607\% &	98.207\% &	98.714\% &	99.015\% \\ 
13 & 81.078\% &	88.993\% &	92.963\% &	95.404\% &	96.924\% &	97.737\% &	98.205\% &	98.603\% \\ 
14 & 50.182\% &	61.804\% &	69.956\% &	76.252\% &	81.232\% &	84.106\% &	86.089\% &	88.310\% \\ 
15 & 57.185\% &	75.790\% &	84.472\% &	89.143\% &	92.066\% &	93.751\% &	95.030\% &	96.128\% \\ 
16 & 82.894\% &	92.270\% &	95.445\% &	97.470\% &	98.611\% &	99.236\% &	99.580\% &	99.735\% \\ 
17 & 53.081\% &	70.204\% &	79.865\% &	86.354\% &	90.656\% &	92.955\% &	94.902\% &	96.052\% \\ 

\hline
\end{tabular}
}
\caption{Successful rate of Technique \text{II} after improvement (Experiments)}
\label{tbl:SR}
\end{table}

\end{document}